%% file: paper.tex
\documentclass[sigconf, anonymous=false]{acmart}
\usepackage[ruled,vlined]{algorithm2e}
\usepackage{array}
\usepackage{multirow}
\usepackage{float}
\usepackage{balance}
\usepackage{subfigure}
\usepackage{booktabs}
\AtBeginDocument{%
  \providecommand\BibTeX{{%
    \normalfont B\kern-0.5em{\scshape i\kern-0.25em b}\kern-0.8em\TeX}}}

\copyrightyear{2021} 
\acmYear{2021} 
\setcopyright{acmcopyright}\acmConference[FPGA '21]{Proceedings of the 2021 ACM/SIGDA International Symposium on Field Programmable Gate Arrays}{February 28-March 2, 2021}{Virtual Event, USA}
\acmBooktitle{Proceedings of the 2021 ACM/SIGDA International Symposium on Field Programmable Gate Arrays (FPGA '21), February 28-March 2, 2021, Virtual Event, USA}
\acmPrice{15.00}
\acmDOI{10.1145/3431920.3439303}
\acmISBN{978-1-4503-8218-2/21/02}

\begin{document}
\fancyhead{}


\author{Thomas Luinaud}
\email{thomas.luinaud@polymtl.ca}
\orcid{0000-0003-4114-1179}
\affiliation{%
  \institution{Polytechnique Montréal}
  \streetaddress{}
  \city{Montréal}
  \state{}
  \country{Canada}
  \postcode{}
}
\author{Jeferson Santiago da Silva}
\email{jeda@kaloom.com}
\affiliation{%
  \institution{Kaloom Inc.}
  \streetaddress{}
  \city{Montréal}
  \state{}
  \country{Canada}
  \postcode{}
}
\author{J.M. Pierre Langlois} 
\email{pierre.langlois@polymtl.ca}
\affiliation{%
  \institution{Polytechnique Montréal}
  \streetaddress{}
  \city{Montréal}
  \state{}
  \country{Canada}
  \postcode{}
}
\author{Yvon Savaria}
\email{yvon.savaria@polymtl.ca}
\affiliation{%
  \institution{Polytechnique Montréal}
  \streetaddress{}
  \city{Montréal}
  \state{}
  \country{Canada}
  \postcode{}
}

\renewcommand{\shortauthors}{Thomas Luinaud et al.}


\title{Design Principles for Packet Deparsers on FPGAs}

\begin{abstract}
  \input{abstract.tex}
\end{abstract}

\maketitle
\input{content.tex}

\bibliographystyle{ACM-Reference-Format}
\balance
\bibliography{ref}

\end{document}

%% file: abstract.tex
The P4 language has drastically changed the networking field as it allows to quickly describe and implement new networking applications.
Although a large variety of applications can be described with the P4 language, current programmable switch architectures impose significant constraints on P4 programs. 
To address this shortcoming, FPGAs have been explored as potential targets for P4 applications. 
P4 applications are described using three abstractions: a packet parser, match-action tables, and a packet deparser, which reassembles the output packet with the result of the match-action tables. 
While implementations of packet parsers and match-action tables on FPGAs have been widely covered in the literature, no general design principles have been presented for the packet deparser. 
Indeed, implementing a high-speed and efficient deparser on FPGAs remains an open issue  because it requires a large amount of interconnections and the architecture must be tailored to a P4 program. 
As a result, in several works where a P4 application is implemented on FPGAs, the deparser  consumes a significant proportion of chip resources. 
Hence, in this paper, we address this issue by presenting design principles for efficient and high-speed deparsers on FPGAs. 
As an artifact, we introduce a tool that generates an efficient vendor-agnostic deparser architecture from a P4 program.
Our design has been validated and simulated with a cocotb-based framework.
The resulting architecture is implemented on Xilinx Ultrascale+ FPGAs and supports a throughput of more than 200 Gbps while reducing resource usage by almost 10$\times$ compared to other solutions.

%% file: content.tex
\section{Introduction}
The P4 Domain Specific Language (DSL)~\cite{10.1145/2656877.2656890} has reshaped the networking domain as it allows describing custom packet forwarding applications with much flexibility.
There has been a growing interest in using FPGAs to offload networking tasks.
For instance, Microsoft already deploy FPGAs in their  data centers to implement the data plane of Azure servers~\cite{cloud_scale_acceleration_architecture}.
In-network computing is another avenue where FPGAs have recently been considered~\cite{10.1145/3302424.3303979}. 
In addition, several recent works  exploit FPGA reconfigurability to create programmable data planes and implement P4 applications~\cite{10.1145/3050220.3050234, 10.1145/3289602.3293924,BENACEK201822}.

\begin{figure}
  \centering
  \includegraphics[width=\columnwidth]{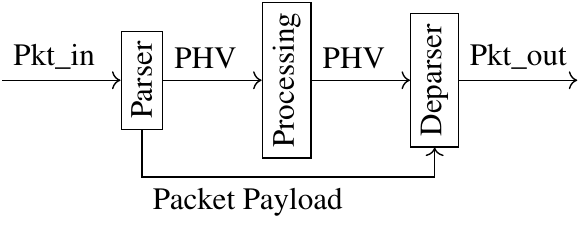}
  \Description[A diagram showing the implementation of a P4 program]{}
  \caption{Considered switch architecture}
  \label{fig:P4Impl}
\end{figure}

As presented in Figure~\ref{fig:P4Impl}, a P4 application comprises three abstractions: the packet parser, the processing stage (match-action tables), and the packet deparser~(\S\ref{sec:P4ProgramImpl}).
While designs of efficient packet parsers on FPGA have been widely explored~\cite{10.1145/3174243.3174270, BENACEK201822, Attig:2011:GPP:2065093.2065215}, little effort has been dedicated to the implementation of efficient packet deparsers. 
First, to the best of our knowledge, only a single paper covers this topic~\cite{8735524}. 
However,~\citeauthor{8735524}~\cite{8735524} report only the FPGA resource consumption for a 100   Gbps packet deparser, while the design principles and microarchitectural details are not covered. Second, as \citeauthor{9098978}~\cite{9098978} have observed, a packet deparser can consume more than 80\% of the resources needed to implement a complete pipeline, which can jeopardize the ability of FPGAs to implement more complex P4 applications.

This paper introduces an open-source solution to generate efficient and high-speed packet deparsers on FPGAs. 
It lays the foundations for the design principles of packet deparsers on FPGAs. 
It comprises an architecture and a compiler to generate a deparser from a P4 program.

The deparser compiler~(\S\ref{sec:depComp}) is described in Python and generates synthesizable VHDL code for the proposed deparser architecture.
The generated architecture leverages the inherent configurability of FPGAs to avoid hardware constructs that cannot be efficiently implemented on FPGAs, such as crossbars or barrel shifters~\cite{10.1145/1086297.1086325, 6412118}. 

The simulation environment is based on \textit{cocotb}~\cite{cocotb}, which allows using several off-the-shelf Python packages, such as Scapy, to generate test cases.
In addition, it is possible to connect the design under test with virtual network interfaces~\cite{cocotb-ping}.
As a result, behavioural validation can be done using the P4 behavioural reference model~\cite{p4BMv2}.

The generated architecture has been evaluated for a variety of packet headers.
The evaluations show that the generated deparser supports more than 200 Gbps packet throughput while reducing the resource usage by more than 10$\times$ compared to state-of-the-art solutions~(\S\ref{sec:results}).

The contributions of this paper are as follows:
\begin{itemize}
\item A deparser architecture that leverages FPGA configurability~(\S\ref{sec:depArch});
\item An open-source P4-to-VHDL packet deparser compiler~(\S\ref{sec:depComp}); and,
\item A simulation environment based on \textit{cocotb} to simplify deparser verification~(\S\ref{sec:results}).
\end{itemize}

\section{Packet Processing}
This section introduces the P4 language and how P4 program components are organized to describe packet processing.

\subsection{P4 language}
P4~\cite{10.1145/2656877.2656890} is an imperative DSL used to describe custom packet processing on programmable data planes.

\subsubsection{Overview of P4 programs}
There are four components that structure a P4 program: \verb|header|, \verb|parser|, \verb|control|, and \verb|switch|. 
A \verb|header| is a structure composed of fields of specific width and a \emph{validity bit}. 
A struct of headers is used to define the set of headers that can be processed by a P4 program.
The \verb|parser| block expresses in which order and how to extract packet headers.
The \verb|Control| block describes the operations to perform on headers.

\subsubsection{Control Operations}
In a control block, multiple operation types can be performed to modify headers.
Two specific operations are of interest for the deparser, \verb|setValid| and \verb|setInvalid|, which can be used to set a header validity bit to valid or invalid, respectively.

In P4, control blocks also implement the deparsing logic.
These blocks are composed of a series of \verb|emit| statements.
First, the order of these statements determines in which order headers are emitted. 
Second, a header is only emitted when its \emph{validity bit} is set.

Because the sequence of \verb|emit| statements determines the header emission order, and because the  \emph{validity bit} can be altered by previous control blocks, the deparser must be able to insert or remove headers at runtime.

\subsection{P4 Program Components}\label{sec:P4ProgramImpl}
This paper considers the switch structure proposed by \citeauthor{BENACEK201822}~\cite{BENACEK201822}, composed of three parts: a parser, a processing part and a deparser, as presented in Figure~\ref{fig:P4Impl}.

\paragraph{Parser}
The parser takes as input a packet and generates a Packet Header Vector (PHV) and a packet payload.
In our design, we assume that the PHV is composed of two parts : the \textit{PHV\_data} bus containing header data and a bitmap vector, the \textit{PHV\_valid} bus, indicating the \emph{validity bit} of each header component.
We also assume that the packet payload is sent through a streaming bus with the first byte at position 0.

\paragraph{Processing}
The processing part takes as input the PHV from the parser and outputs a modified PHV, which is forwarded to the deparser.
The operations on the PHV can either be header data modifications or header \emph{validity bit} alteration.

\paragraph{Deparser}
The deparser block takes as input the PHV from the processing part and the payload from the parser.
It outputs the packet to be sent on a streaming bus.

\section{Deparser Architecture Principles}\label{sec:depArch}
In this section we cover the deparser architecture principles. 
First, we introduce a deparser abstract machine. 
Second, we cover the deparser I/O signals. 
Third, we present the microarchitecture of the proposed deparser.
All our design choices use the inherent configurability of the FPGAs and provide configurable blocks for the deparser compiler.

\subsection{Deparser Abstract Machine}
\begin{figure}
  \centering
  \includegraphics[width=.9\columnwidth]{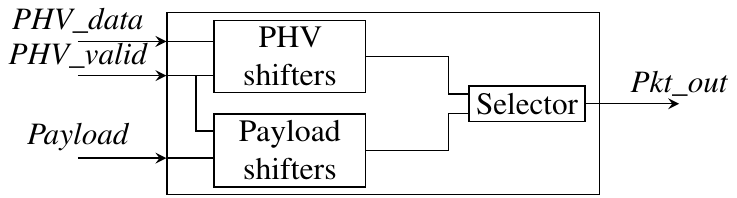}
  \caption{Deparser architecture overview}
  \Description[Deparser overview]{}
  \label{fig:DepDetails}
\end{figure}

The deparser abstract machine is described in Figure~\ref{fig:DepDetails}.
In this architecture, we assume that the PHV is buffered and arrives at the deparser together with a PHV valid vector.

Algorithm~\ref{alg:phvShifter} presents the pseudo-code for the PHV shifters module while Algorithm~\ref{alg:payloadShifter} illustrates the payload shifter.

\begin{algorithm}
    \caption{PHV shifter}\label{alg:phvShifter}
    \SetAlgoLined
    \SetKwInOut{Input}{input}
    \SetKwInOut{Output}{output}
    \Input{$phv$: The PHV}
    \Input{$phv\_valid$: PHV valid vector}
    \Output{$phv\_aligned\_bus$: The PHV aligned bus}
    $pos$ $\leftarrow$ 0\;
    \ForEach{valid in phv\_valid}
    {
		\If{$ph$.\texttt{isValid()}}{
			$phv\_aligned\_bus$.\texttt{insert}($pos$, $phv$.\texttt{at}($pos$, $pos$ + $ph$.\texttt{size()})) \;
			$pos$ $\leftarrow$ $pos$ + $ph$.\texttt{size()}\;
		}
    }
    \KwRet{$phv\_aligned\_bus$}
\end{algorithm}

\begin{algorithm}
    \caption{Payload shifter}\label{alg:payloadShifter}
    \SetAlgoLined
    \SetKwInOut{Input}{input}
    \SetKwInOut{Output}{output}
    \Input{$payload$}
    \Input{$phv\_valid$: PHV valid vector}
    \Output{$payload\_aligned\_bus$: The payload aligned bus}
	$payload\_aligned\_bus$.\texttt{insert}(\texttt{sum\_sizes}($phv\_valid$), $payload$) \;
    \KwRet{$payload\_aligned\_bus$}
\end{algorithm}

The main limiting factors to implement a deparser on FPGAs are the high number of interconnections and barrel shifters required for header insertion.
To limit the use of these blocks, we construct a new packet based on the header and the payload.
Thus, as the P4 deparsing logic can entirely be inferred at compile time and since FPGAs are reconfigurable, we tailor the deparser architecture to a given P4 program in order to alleviate those limiting factors.

We now cover the deparser inputs and outputs.

\subsection{Inputs and Outputs}
The deparser has three inputs and one output.
The output \textit{Pkt\_out} and the input \textit{Payload} are AXI4-stream buses~\cite{amba4axi4}.
The data width of those two buses is a compile time parameter.
The two inputs \textit{PHV\_data} and \textit{PHV\_valid} respectively contain the headers data and validity bits to deparse.
The width of \textit{PHV\_data} and \textit{PHV\_valid} are determined when compiling the P4 application.

\subsection{Microarchitectural Details}
Internally, the deparser is built around three blocks: \texttt{PHV shifters}, \texttt{Payload shifters}, and \texttt{selector}. 
The \texttt{PHV shifters} gets \textit{PHV\_data} and \textit{PHV\_valid} as input, and outputs a frame of headers to emit. 
The \texttt{Payload shifters} receives \textit{Payload} and \textit{PHV\_valid} as input, and generates payload data frames.
Both, \texttt{Payload shifters} and \texttt{PHV shifters} are inputs to the \texttt{selector}. 
The \texttt{Selector} generates the \textit{Pkt\_out} frames according to the status received by \texttt{Payload shifters} and \texttt{PHV shifters}.

\subsubsection{The PHV Shifter}\label{sec:headerShifter}

\begin{figure}
  \centering
  \includegraphics[width=.9\columnwidth]{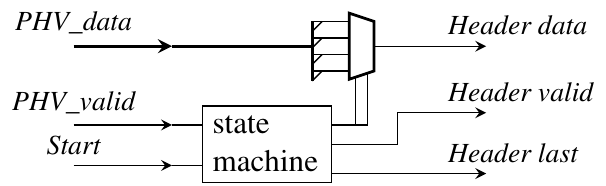}
  \Description[PHV shifter]{}
  \caption{Header shifter for 1 bit}
  \label{fig:phv_shifter}
\end{figure}

The \texttt{PHV shifters} shifts the PHV bits to build the packet.
It is composed  of header shifters presented in Figure~\ref{fig:phv_shifter}.
The maximum number of header shifters is equal to \textit{Pkt\_out} bus width.

The header shifter has three inputs: \emph{PHV\_data}, \emph{PHV\_valid} and a \emph{Start} signal.
It outputs: the \emph{header data}, the \emph{header valid}, and the \emph{header last}.
The \emph{PHV\_valid} and \emph{Start} inputs are connected to the \emph{state machine} module.
The \emph{state machine} module drives the \emph{header valid} and \emph{header last} output.
The \emph{PHV\_data} input is connected to a multiplexer that drives the \emph{header data} output.
The multiplexer selects one of the bits of \emph{PHV\_data} based on one output of the \emph{state machine} module.
The \emph{state machine} is derived from the deparser graph~(\S\ref{sec:depComp}), as well as the number of inputs for the multiplexer.

\subsubsection{The Payload Shifter}\label{sec:payloadShifterArch}
\begin{figure}
  \centering
  \includegraphics[width=.9\columnwidth]{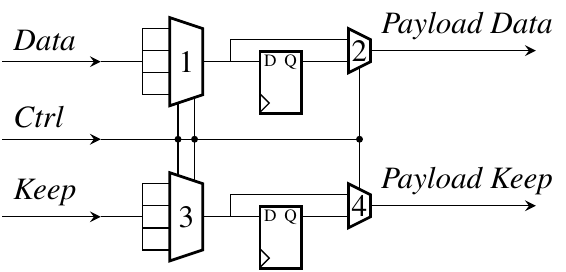}
  \caption{Payload shifter for 1 bit of data}
  \Description[Payload shifter]{}
  \label{fig:payloadShifter}
\end{figure}

The \texttt{Payload shifters} aligns the payload with the emitted headers. 
The basic block of the \texttt{Payload shifters} is shown in Figure~\ref{fig:payloadShifter}. 
It takes \textit{Data}, \textit{Ctrl}, and \textit{Keep} as inputs, and outputs  \emph{Payload Data} and \emph{Payload Keep} signals. 
The bus \textit{Data} and \textit{Keep} are respectively connected to the AXI \texttt{tdata} and \texttt{tkeep} signals of the deparser's \textit{Payload} input buses. 
Each byte of this bus is connected to one input of multiplexer 1 in Figure~\ref{fig:payloadShifter}. 
Each bit of the \textit{Keep} signal is connected to one input of the multiplexer 3. 
The \textit{Ctrl} signal determines which input of multiplexers 1 and 3 should be selected.
Finally, the output of multiplexers 1 and 3 can be registered to delay the data output by one cycle.
Multiplexers 2 and 4 select either the current value or the delayed one.
The value of the \textit{Ctrl} bus is set by a small and constant associative memory generated at compile time.

\subsubsection{The Selector}
\begin{figure}
  \centering
  \includegraphics[width=.9\columnwidth]{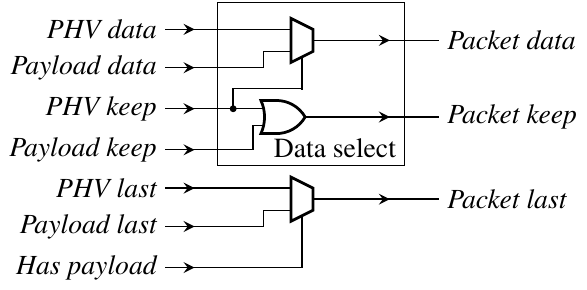}
  \caption{Deparser selector for 1 bit of data}
  \Description[selector]{}
  \label{fig:selector}
\end{figure}

This block selects the right output data between \texttt{Payload shifters} and \texttt{PHV shifters} and generates the AXI4-stream output signals \emph{Packet data}, \emph{Packet keep} and \emph{Packet last}.
The selector takes as input the output of the \texttt{Payload shifters} and the output of the \texttt{PHV shifters} as shown in Figure~\ref{fig:selector}.

The \emph{Packet data} and \emph{Packet keep} signals are assigned by the block \emph{Data Select}.
This block is duplicated to assign all the packet data bits.
The \emph{Packet last} signal is either \emph{PHV last} or \emph{Payload last} according to the presence of a payload indicated by the input signal \emph{Has payload}.

\subsubsection{Multiplexers on FPGAs}
The different presented building blocks are highly dependent on the multiplexer implementation on FPGAs.
We chose to use multiplexers since they are efficiently implemented on FPGAs.
Indeed, a 16:1 multiplexer consumes a single slice on a Xilinx FPGA~\cite{xilinxVirtex7}.
While having a very large multiplexer can become expensive, we know that the number of inputs for each multiplexer will be reduced to a minimum by the compiler~(\S\ref{sec:subDagTranslation}).

\section{Deparser generation}\label{sec:depComp}
The deparser can be represented as a Directed Acyclic Graph (DAG).
To generate a deparser DAG, a P4 program is compiled into a JSON file using the \verb|p4c-bm2-ss| compiler~\cite{P4c}.
The generated JSON file is then used to generate a deparser DAG.
It is possible to optimize this DAG, but this optimization was left to future works.
The rest of this section presents the process of generating the different deparser modules from a deparser DAG.

\subsection{The Deparser DAG}
\begin{figure}
    \includegraphics[width=\linewidth]{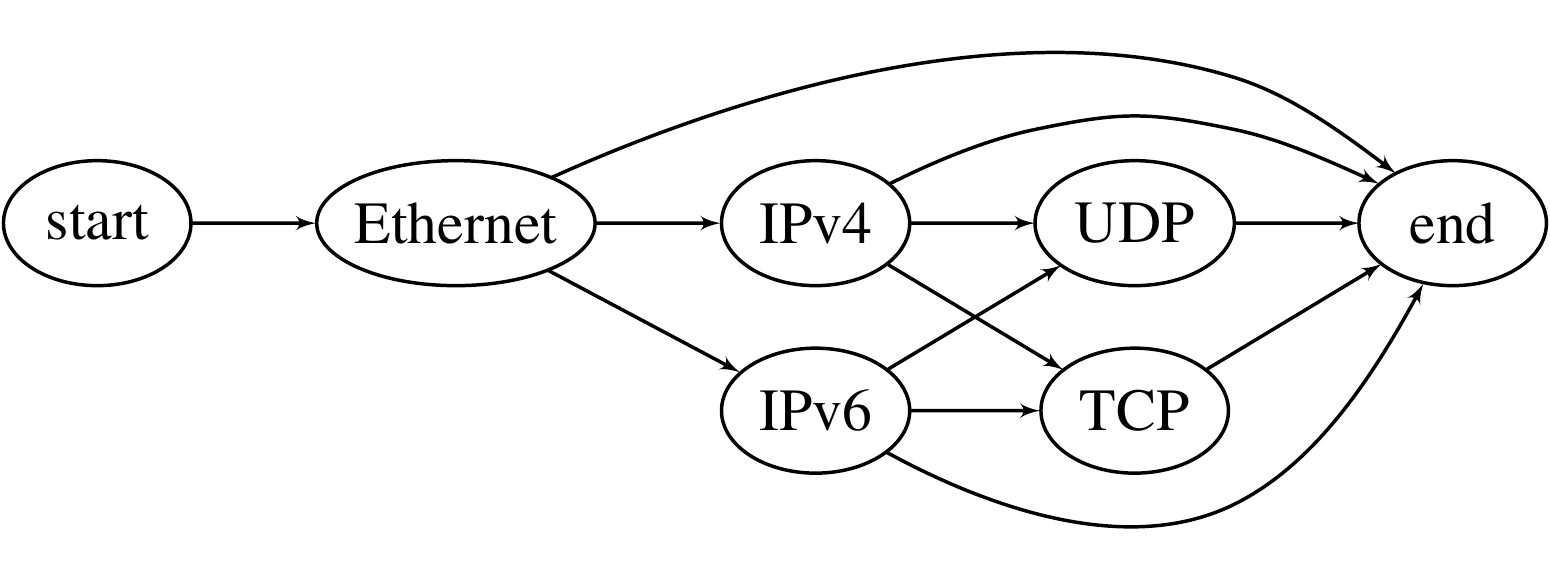}
    \caption{A possible deparser DAG with Ethernet, IPv4, IPv6, TCP and UDP headers}
    \label{fig:graphExample}
\end{figure}

An example of a deparser DAG, for Ethernet, IPv4, IPv6, TCP and UDP packets, is presented in Figure~\ref{fig:graphExample}.
Each node of the DAG, excluding the \emph{start} and the \emph{end}, represents a header.
Each arrow of the DAG indicates the possible next headers to emit.
Each path between \emph{start} and \emph{end} represents a possible set of headers to emit. The list of all possible paths is given in Table~\ref{tab:possiblePath}.

\begin{table}
    \caption{Possible sequences of headers based on Figure~\ref{fig:graphExample} with total header size~\subref{subtab:headerSeq} and the size of each headers~\subref{subtab:headerSize}}
    \label{tab:possiblePath}
    \centering
    \subtable[]{
    \label{subtab:headerSeq}
    \begin{tabular}{@{}lm{1cm}<{\centering}@{}}
    \toprule
    Path & Size (Bytes) \\
    \midrule
    Ethernet & 14 \\
    Ethernet->IPv4 & 34 \\
    Ethernet->IPv4->TCP & 54 \\
    Ethernet->IPv4->UDP & 42\\
    Ethernet->IPv6 & 54\\
    Ethernet->IPv6->TCP & 74\\
    Ethernet->IPv6->UDP & 62\\
    \bottomrule
    \end{tabular}}\hfill
    \subtable[]{\label{subtab:headerSize}
    \begin{tabular}{@{}lm{1cm}<{\centering}@{}}
    \toprule
    Header & Size (Bytes) \\
    \midrule
    Ethernet & 14 \\
    IPv4 & 20 \\
    IPv6 & 40 \\
    TCP & 20 \\
    UDP & 8 \\
    \bottomrule
    \end{tabular}}
\end{table}

There are two parts in order to obtain a deparser from the DAG.
The first part transform the deparser graph to generate the \texttt{PHV shifters}.
The second part uses the deparser graph to generate the \texttt{Payload shifters}.

\subsection{PHV Shifters Generation}
To generate the header shifters of the \texttt{PHV shifters}, we build sub-DAGs of the deparser graph.
Each sub-DAG represents one header shifter block~(\S\ref{sec:headerShifter}).
Since most network protocols are byte aligned~\cite{BENACEK201822}, we build one sub-DAG per output byte.
This allows merging the PHV shifter state machines, hence, reduce the deparser architectural complexity.

\subsubsection{Sub-DAGs Generation}
In a sub-DAG, every node contains the header and the byte to extract.
Each edge indicates the header validity condition to go to the corresponding next node.
We propose Algorithm~\ref{alg:subDAGgEn} to generate sub-DAGs.

\begin{figure}
    \centering
    \includegraphics[width=\columnwidth]{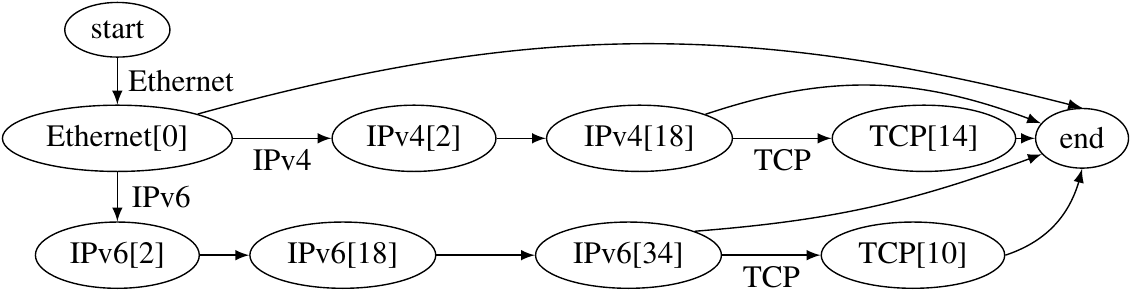}
    \caption{Sub-DAG for byte 0 from Figure~\ref{fig:graphExample} DAG}
    \label{fig:subDag}
\end{figure}

The proposed algorithm goes through all the possible sequences of header emissions by traversing the deparser graph.
We assign each byte of a sequence to a sub-DAG.
When it is the first time a sub-DAG processes a byte from a specific header, we set the edge condition to this header.
Figure~\ref{fig:subDag} shows a sub-DAG generated with Algorithm~\ref{alg:subDAGgEn} using the deparser DAG of Figure~\ref{fig:graphExample} and a 128-bit output bus.

\subsubsection{Sub-DAG Translation}\label{sec:subDagTranslation}
A sub-DAG translation is decomposed in two parts: the header shifter mux generation and the state machine generation.
The number of inputs of the generated mux is equal to the number of nodes in the sub-DAG.
The state machine is derived from the sub-DAG where each node represents a state and each edge a transition.
The header byte position is converted into an input position of the multiplexer.

\begin{algorithm}
    \caption{Deparser Sub-DAGs generation}\label{alg:subDAGgEn}
    \SetAlgoLined
    \SetKwInOut{Input}{input}
    \SetKwInOut{Output}{output}
    \Input{A deparser DAG $G_d$, $w$ number of sub-DAG}
    \Output{$w$ sub graph}
    $Sg$ $\leftarrow$ list of $w$ empty-DAG\;
    \ForEach{path $p$ between \emph{start} and \emph{end}  in $G_d$}
    {
        $Sgi \leftarrow 0$ \tcp*[r]{SubGraph indice}
        prev\_node $\leftarrow$ [] \;
        \lFor{$i \leftarrow 0$ \KwTo $w$}{prev\_node.append(None)}
        \ForEach{header $h$ in $p$}{
            \For{$i \leftarrow 0$ \KwTo nbByte($h$)}
            {
                $newNode$ $\leftarrow$ node extract $h$ position $i$ \;
                $Sg_{Sgi}$.insert($newNode$) \;
                $newEdge$ $\leftarrow$ edge prev\_node $ \rightarrow newNode$ \;
                \If(\tcp*[f]{conditional edge}){$Sgi < i$}{
                    $newEdge$.set\_transition\_condition($h$) \;
                }
                $Sg_{Sgi}$.insert($newEdge$) \;
                prev\_node[$Sgi$] $\leftarrow newNode$ \; 
                $Sgi \leftarrow (Sgi + 1)$ mod $w$ \tcp*[h]{next sub-DAG}
            }
        }
    }
    \KwRet{$Sg$}
\end{algorithm}

\subsection{Payload Shifter Associative Memory Creation}
The payload shifter architecture is presented in~(\S\ref{sec:payloadShifterArch}).
We use the graph to generate the associative memory that drives the \emph{Ctrl} signal.
This memory is generated in two steps using the deparser graph.
First, we determine the set $\text{PH}$ of possible valid headers by looking at all the possible paths between \emph{start} and \emph{end} in the deparser graph.
Each element  $\text{ph}$ in the set $\text{PH}$ is composed of the \emph{PHV\_valid} bus value and the sum $ph_w$ of all headers widths.

Once the set $\text{PH}$ is built, for each possible element $\text{ph} \in \text{PH}$, we assign a value to \emph{Ctrl}.
For each possible \emph{PHV\_valid} value, we calculate the offset for the payload.
The offset is obtained using the equation $\text{Offset} = ph_{w} \pmod w$, where $w$ is the bus width and $PH_w$ represents the total number of bits of the emitted header.
Finally, we set the values of the \emph{Ctrl} bit connected to the multiplexers 2 and 4 of Figure~\ref{fig:payloadShifter} for each payload shifter positioned below the offset value.

\begin{table*}
  \caption{Synthesis results of deparser DAG on a Xilinx xcvu3p FPGA}
  \label{tab:resultImpl}
  \begin{tabular}{@{}lcccccccccc@{}}
    \toprule
\multirow{2}{*}{Test} & \multirow{2}{0.8cm}{\centering width (bits)} & \multirow{2}{2cm}{\centering worst latency (cycles)} & \multicolumn{4}{c}{Deparser DAG} & \multicolumn{4}{c}{Parser DAG}\\
    & & & LUTs & FFs & BRAMs & Frequency & LUTs & FFs & BRAMs & Frequency\\
    \midrule
    \multirow{4}{*}{\textbf{T1}} & 64 & 19 & 1517 & 402 & 0 &  448 MHz & 630 & 395 & 0 &    448 MHz     \\ 
                                 & 128 & 13 & 2066 & 784 & 0 &  448 MHz & 1019 & 782 & 0 &   448 MHz  \\
                                 & 256 & 10 & 2862 & 1522 & 0 & 448 MHz & 1722 & 1365 & 0 &  448 MHz \\
                                 & 512 & 8 & 9127 & 3002 & 0 & 469 MHz & 5777 & 2696 & 0 &  469 MHz \\ 
    \midrule
    \multirow{4}{*}{\textbf{T2}} & 64  & 20 & 1798 & 404 & 0 &   448 MHz & 746 & 402 & 0   &  448 MHz    \\ 
                                 & 128 & 13 & 2664 & 808 & 0 &   448 MHz & 1075 & 758 & 0 &   448  MHz     \\
                                 & 256 & 10 & 4879 & 1602 & 0 &  469 MHz & 1667 & 1370 & 0 &  469  MHz   \\
                                 & 512 & 8 & 11212 & 3197 & 0 & 448 MHz & 5811 & 2691 & 0 &  448  MHz  \\
    \midrule                                                                  
    \multirow{4}{*}{\textbf{T3}} & 64  & 22 & 2137 & 390 & 6 &    448 MHz & 1039 & 372 & 2 &   448  MHz    \\ 
                                 & 128 & 14 & 3962 & 780 & 14 &   448 MHz & 1858 & 758 & 5 &   448  MHz   \\
                                 & 256 & 10 & 6598 & 1525 & 32 &  448 MHz & 3842 & 1450 & 29 & 448 MHz \\
                                 & 512 & 8 & 14287 & 3116 & 41 & 469 MHz & 8603 & 2931 & 32 & 469 MHz  \\
    \bottomrule
  \end{tabular}
\end{table*}

\begin{table}
  \caption{Implementation results for a 512 bits output data bus compared with previous work}
  \label{tab:resultComp}
  \begin{tabular}{@{}lcccccm{1.7cm}<{\centering}@{}}
    \toprule
    Test & Work & Slice & LUT & FF & BRAM & Throughput (Gbps) \\
    \midrule
    \multirow{3}{*}{\textbf{T1}} & Our & 3144 & 9 k & 3 k &  0 & 200  \\ 
                                 & SDNet & N/A & 77 k & 95 k & 116.5 & 160 \\
                                 & SDNet & N/A & 78 k & 95 k & 116.5 & 256 \\
    \midrule
    \multirow{3}{*}{\textbf{T2}} & Our & 3922 & 11.2 k & 3.2 k & 0 & 220 \\
                                 & SDNet & N/A & 98 k & 119 k & 149.5 & 240 \\
                                 & \cite{BENACEK201822} & 20 k & N/A & N/A & N/A & 120 \\ 
    \midrule
    \multirow{3}{*}{\textbf{T3}} & Our & 4770 & 14 k & 3 k & 20.5 & 140 \\
                                 & SDNet & N/A & 137 k & 165 k & 229.5 & 160 \\
                                 & SDNet & N/A & 139 k & 165 k & 229.5 & 220 \\
                                 & \cite{BENACEK201822} & 24 k & N/A  & N/A & N/A & 120\\ 
    \bottomrule
  \end{tabular}
\end{table}

\section{Results}\label{sec:results}

This section presents the results of this work.
First, we describe the experimental setup.
Then we present the impact of the compiler parameters on the generated architecture.
Finally, we present and compare the implementation results with previous work.

\subsection{Experimental Setup}

We have generated deparsers for the following three protocol stacks:
\begin{itemize}
\item \textbf{T1} : Ethernet, IPv4/IPv6, TCP/UDP
\item \textbf{T2} : Ethernet, IPv4/IPv6, TCP/UDP, ICMP/ICMPv6
\item \textbf{T3} : Ethernet, 2 $\times$ VLAN, 2 $\times$ MPLS, IPv4/IPv6, TCP/UDP, ICMP/ICMPv6
\end{itemize}

To validate our work, we developed a simulation platform based on the \textit{cocotb} framework~\cite{cocotb}.
We developed \textit{cocotb} drivers and monitors for the AXI4-stream bus, allowing us to rapidly evaluate different deparser configurations.
Xilinx Vivado 2019.1 was used for synthesis and place-and-route.
To allow reproducibility, our codes are open\footnote{\url{https://github.com/luinaudt/deparser/tree/FPGA_paper}}.

\subsection{Impact of Compiler Parameters}
To evaluate the impact of the graph complexity, we generated and synthesized deparsers from both non-optimized deparser DAGs and parser DAGs considered as optimized deparser DAGs.
Using the parser DAG as a deparser DAG was the proposed implementation in P4\textsubscript{14}~\cite{p414}.
The Block RAM (BRAM), Look Up Table (LUT), and Flip Flop (FF) usage for each synthesis run when targeting a Xilinx xcvu3p-3 FPGA are presented in Table~\ref{tab:resultImpl}.
The results indicate that two factors dominate resource usage: graph complexity and data bus width.

\paragraph{Graph complexity}
The graph complexity is impacted by the deparser code and the extent to which the graph was simplified.
Since simplified deparser DAGs have fewer edges, the state machine for their PHV\_shifter and the size of the associative memory for the payload shifter are reduced.
In addition, fewer nodes are required for each sub-DAG.
As a result, the number of inputs for the PHV shifter mux is reduced.
For example, in \textbf{T1}, there are 5 headers.
This results into a total of 32 paths for a non-optimized deparser DAG, while the simplified parser graph contains only 7 paths.

\paragraph{Bus Width}
In addition to the graph complexity, the bus width impacts resource consumption.
The proposed design has a latency of 6 clock cycles.
Also, the latency to output a packet is a function of the total header length to emit and the bus width.
As presented in Table~\ref{tab:resultImpl}, for wider buses, the worst-case latency is reduced compared to smaller buses.
The worst-case latency for header emission can be calculated with the following equation: 
\begin{equation}
    \text{latency} = \left\lceil \frac{\text{total header length}}{\text{bus width}} \right\rceil + 6
\end{equation}

Also, increasing the bus width increases LUTs and FFs usage.
For data buses varying from 64 to 256 bits, there is a slight increase in resource usage, however, this increase becomes significant for a 512-bit bus.
Two factors can explain this higher complexity. 
First, the minimum number of multiplexer increases at a rate of 1 multiplexer per output bit. 
Second, with larger buses, due to header alignment, more headers can be appended to the bus for each output frame. Hence, there is less reuse of possible inputs for the PHV shifter.

\subsection{Implementation Results}
We also implemented non-optimized deparser DAGs for the three protocol stacks with a data bus of 512 bits.
We compared the deparser implementation results against the deparser generated by the Xilinx SDNet 2017.4~\cite{XilinxSDNET} and  \citeauthor{BENACEK201822}~\cite{BENACEK201822}  deparsers.
The results of these implementations are shown in Table~\ref{tab:resultComp}.
Our deparser supports a throughput 20 Gpbs greater than the deparser proposed by \citeauthor{BENACEK201822}~\cite{BENACEK201822}, while reducing by $5 \times$ the resource usage.
Compared to the parser generated by Xilinx SDNet~\cite{XilinxSDNET}, in the worst case, our deparser supports a throughput that is 60 Gbps lower, but our deparser uses almost $10 \times$ fewer resources. 

When comparing implementation and synthesis results, the resource consumption remains stable.
In the case of \textbf{T1} and \textbf{T2}, the performance after place-and-route is almost the same. However, the maximum clock frequency could be improved by pipelining the multiplexers, without significantly impacting resource consumption.
Indeed, the generated architecture consumes less than one FF per slice while a typical slice posses eight FFs.
As a result, unused FFs could be used to pipeline multiplexers, as they would be unlikely to be driven by other modules.

\section{Related Work}\label{sec:related}

\citeauthor{p4fpga} proposed P4FPGA~\cite{p4fpga}.
P4FPGA is an open-source and vendor-agnostic P4-to-FPGA compiler targeting mid-performance FPGAs (10  Gbps).
\citeauthor{p4tosdnet}~\cite{p4tosdnet} proposed integrating Xilinx SDNet P4 compiler~\cite{XilinxSDNET} to the off-the-shelf NetFPGA board~\cite{netfpga}. 
This work exposes the SDNet limitation in implementing the deparser logic, which turned out to be the module with the largest resource consumption of the generated pipeline~\cite{9098978}.
Indeed, as per our experiments, we observed that Xilinx SDNet was unable to optimize unreachable paths in the deparsing graph.

\citeauthor{BENACEK201822}~\cite{BENACEK201822} presented automatic generation of P4-based packet parsers and deparsers to VHDL.
This work extends to deparsers their previous research on packet parsers~\cite{7544769}. However, the deparser architecture and design principles have not been covered because the optimizations were derived from the design of packet parsers.
Other packet parsers research include~\cite{6665172,10.1145/3174243.3174270, Attig:2011:GPP:2065093.2065215}.
\citeauthor{6665172}~\cite{6665172} introduced general design principles for packet parsers, but does not cover the case of packet deparsers.
In addition, \citeauthor{Attig:2011:GPP:2065093.2065215}~\cite{Attig:2011:GPP:2065093.2065215} proposed a language to represent parsers with an architecture and a compiler to implement them on an FPGA. 
Also, \citeauthor{10.1145/3174243.3174270}~\cite{10.1145/3174243.3174270} proposed to use graph optimizations, similarly to our work, to simplify the parser pipeline. 

\section{Conclusion}
P4 has changed the networking landscape as it allows expressing custom packet processing.
In recent years, several works have mapped P4 programs to FPGAs.
However, the majority of these works have focused on implementing packet parsers or match-action stages.
To date, no general packet deparsing principles on FPGAs have been proposed.
Indeed, the naive approach of previous works on generation of deparsing logic has made hardware implementation of this block very costly on FPGAs.
In this work, we tackle this problem by introducing a set of design principles for implementing packet deparsers on FPGAs.
In our work, we proposed an architecture tightly coupled to the FPGA microarchitecture in order to leverage the FPGA's inherent configurability.
We also demonstrate the importance of deparser graph simplification to reduce resource consumption. 
Our results show that our proposed deparser architecture crosses the 100  Gbps throughput boundary while reducing the resource consumption by one order of magnitude.
Finally, to permit reproducibility, we open-sourced our framework and an integrated simulation environment based on \textit{cocotb}.

\section*{Acknowledgements}
The authors thank Thibaut Stimpfling and Bochra Boughzala from Kaloom for their insightful comments.
This work was supported by Kaloom, Intel, Noviflow, the Natural Sciences and Engineering Research Council of Canada, and Prompt.